\documentclass[onecolumn]{elsart}
\usepackage{amssymb}
\usepackage{amsfonts}
\usepackage{amsmath}

\input{tcilatex}

\begin{document}

\begin{frontmatter}%

\title{Physical significance of the curvature varifold-based description of crack nucleation}%

\author{Paolo Maria Mariano}%

\address{DICeA, University of Florence, \\ via Santa Marta 3, I-50139 Firenze (Italy) \\ e-mail: paolo.mariano@unifi.it}%

\begin{abstract}
The nucleation and/or growth of cracks in elastic-brittle solids has been
recently described in \cite{GMMM} in terms of a special class of measures
and with a variational technique requiring the minimization of a certain
energy over classes of bodies. Here, the physical foundations of the theory
and the basic ideas leading to it are described and commented further on. A
view on certain possible developments and shifts toward different settings
is also given. This article has expository character.%
\end{abstract}%


\end{frontmatter}%

When the adjective `variational' is attributed to models of some physical
phenomena, impulsively one may think that the physical situation under
scrutiny is conservative. Variational means, in fact, that one is managing
some functional, essentially an energy, and is asking that it attains its
minimum on some function space. The situation is well known and is typical,
for example, of the determination of ground states of elastic bodies under
prescribed boundary conditions. The energy is defined on the reference place 
$\mathcal{B}$ occupied by the body in the ambient space, a place fixed once
and for all. Its minimum is required to be attained over a class of
one-to-one orientation preserving maps.

Variational descriptions of the nucleation and the evolution of defects like
cracks however exist. At least in principle, the question of their physical
appropriateness can be posed, due to the dissipative character of the
phenomena they are referred to. Of course, one can say that the quest of
something to be minimized has rather instrumental character because in
nature some economic principle always appears somewhere and the calculus of
variations is at a stage of development that it can be desirable to be under
conditions of using it. This point of view can be an answer. However, the
answer can be even deeper.

When one think of the nucleation of defects, in fact, one manages a \emph{%
mutant body}. Mutations occur, in fact, in the gross shape of the body at
the continuum level, and they can be naturally pictured as mutations of the
reference place $\mathcal{B}$ which is now \emph{not} fixed once and for all
as for example in the standard formulation of elasticity. In principle, one
can imagine of having at disposal a class of possible bodies occupying $%
\mathcal{B}$, every member of the class differing from the others by the
defect pattern. Once boundary conditions are prescribed, a variational
principle can select the resulting body within a class. This way, a
variational approach to the defect nucleation -- the latter imposed by the
interplay between boundary condition and nature of the material -- is
physically significant when it involves a minimization process of the free
energy (for example) over an entire class of possible bodies. This point of
view is the one adopted here and is also the answer to the question of
physical appropriateness of some variational approaches. The energy
dissipated in the nucleation (or growth) process is individuated in the gap
arising from the body in the original shape -- the one existing before the
assignment of boundary conditions -- and the actual shape (i.e. the body
plus the defect pattern) obtained by the minimization procedure. Of course
the explicit evaluation of this gap can be arduous.

This point of view interprets the physical aspects of two existing
variational models of fracture mechanics. Both models are recalled below.
The attention is focused on the second one in which cracks are described
through appropriate measures, the so-called curvature varifolds. The
physical significance of the approach is discussed here. It is shown also
that the point of view can be adopted in other circumstances dealing with
the nucleation of defects stratified over manifolds with different
dimensions.

\section{ \ \ }

Under the suggestion of the pioneering Griffith's approach to fracture in
brittle-elastic solids, a variational model in fracture mechanics has been
proposed in \cite{FM}. It is based on a requirement of minimality of the
overall energy $\mathfrak{E}$\ which accounts for the macroscopic
deformation and the possible presence of cracks, and is defined by 
\begin{equation*}
\mathfrak{E}\left( \mathcal{C},u\right) :=\int_{\mathcal{B}}e\left(
x,Du\left( x\right) \right) \text{ }dx+\int_{\mathcal{C}}\phi \text{ }d%
\mathcal{H}^{2},
\end{equation*}%
where $\mathcal{B}$ is the regular\footnote{%
The type of regularity is specified later.} region occupied in the
three-dimensional ambient space by the body, $Du\left( x\right) $ the
spatial derivative evaluated at $x\in \mathcal{B}$ of the differentiable
transplacement map $x\longmapsto u\left( x\right) \in \mathbb{R}^{3}$ -- the
map defining the actual (deformed) place $u\left( \mathcal{B}\right) $ -- $%
\mathcal{C}$\ the representation in $\mathcal{B}$ of a surface-like crack
occurring in $u\left( \mathcal{B}\right) $, $e$ the elastic energy, $\phi $
a constant surface energy, $d\mathcal{H}^{2}$ the two-dimensional Hausdorff
measure. The map $x\longmapsto u$ is also assumed, as usual, to be
one-to-one outside $\mathcal{C}$, orientation preserving (i.e. $\det
Du\left( x\right) >0$), and such that the global invertibility condition%
\begin{equation*}
\int_{\mathcal{B}}\tilde{f}\left( x,u\left( x\right) \right) \det Du\left(
x\right) \text{ }dx\leq \int_{\mathbb{R}^{3}}\sup_{x\in \mathcal{B}}\tilde{f}%
\left( x,z\right) \text{ }dz,
\end{equation*}%
holds for all $\tilde{f}\in C_{0}^{\infty }(\mathcal{B}\times \mathbb{R}^{3})
$,\ a condition allowing frictionless self-contact of the boundary while
still preventing self-penetration (for details on this last condition see 
\cite{GMS}). In other words, $x\longmapsto u$ is an orientation preserving
homeomorphism outside the subset of $\mathcal{C}$ containing its jump set.

The requirement is then that \emph{at each instant} $t\in \lbrack 0,\bar{t}]$
\emph{of a cracking process the pair} $\left( \mathcal{C},u\right) \left(
t\right) $ \emph{realize a minimum of the global energy} $\mathfrak{E}$\
with $\mathcal{C}$ an admissible crack. Admissibility is intended in the
sense that $\mathcal{C}$ is a rectifiable set (specifically the image of a
countable number of Lipschitz maps) with zero volume measure. The interval
of time is then discretized and minimality is required at time steps.
Minimizers are pairs $\left( \mathcal{C},u\right) $: the minimum problem has
two variables.

Notwithstanding simplicity and elegance of this model, the evaluation of
minima of the energy at each time step involves a number of analytical
problems. The essential difficulty arises in controlling in three dimensions
minimizing sequences of surfaces leading to the possible actual crack, or
better to its picture $\mathcal{C}$ in the reference place. However, when
the entire crack is open, $\mathcal{C}$ coincides with the jump set of the
transplacement field $x\longmapsto u\left( x\right) $. The convenient
simplification of identifying cracks with the jump sets of displacement
fields has been then adopted. Bounded variation (BV) or special bounded
variation (SBV) functions have been then involved as candidates to be
minimizers of the energy considered as a functional of the sole $u$. The
energy is then considered as the one of a simple Cauchy's body\footnote{%
In continuum mechanics a body is considered in primitive sense as a set with
elements called the material elements and identified even vaguely with
molecular or atomic aggregates. The selection of material elements, also
called representative volume elements, is matter of modelling. Once even a
rough idea of them is formulated -- in a sense one is specifying what are
the peculiar physical aspects of the material texture -- the essential step
is to furnish geometrical structure to the body which would be otherwise
just an abstract set. The representation of the interactions is then
straighforward, dual in the sense of power. The geometrical representation
of the material elements is then matter of modelling and can be even
minimalist: in fact, one can choose to assign to every material element only
the place that it occupies in the ambient space. I use to call Cauchy's
bodies those bodies for which the minimalist approach summarized above is
sufficient to represent the essential peculiarities of their morphology, and
the representation of inner actions is just in terms of standard stresses.}
and minimizers are sought in a space of maps including candidates to be
reasonable descriptors of the elastic-brittle behavior. Remind that the
choice of function spaces where one researches minimizers has constitutive
nature. A difficulty has been however encountered: theorems allowing the
selection of fields with discontinuity sets describing reasonable
(physically significant) crack patterns seem to be not available at least in
the current literature (see \cite{BFM} for a pertinent review of the results
along this line; also \cite{DFT}, \cite{FraMil}, \cite{D}). Moreover, this
kind of approach seems to be not able to account for partially opened
cracks. In fact, in this case the transplacement field is continuous across
the closed part of the crack although the material bonds are broken in the
actual place. The description of partially closed cracks can have even
stringent interest in the time-discretized procedure sketched above. In
fact, when such a procedure is applied by taking into account a loading
program described by time-dependent boundary conditions, a program implying
even the growth of pre-existing cracks besides the nucleation of a new
fracture at the $n-$th instant, it can happen that the cracks may close even
partially, and then they re-open at subsequent time steps. Moreover, the
discovery of physically appropriate crack patterns is another key point.

Such questions have been tackled from a different point of view in \cite%
{GMMM} (see also additional comments in \cite{2}, \cite{3}) by using tools
from geometric analysis. The skeletal features of the theory are listed in
the ensuing items.

\begin{description}
\item[(\emph{i})] Distinction is made between cracks and jump set of the
transplacement field, as in \cite{FM}. The latter set is however constrained
to be contained within the crack pattern. Differently from all previous
proposals, the crack pattern is described through measures over a fiber
bundle with typical fiber the Grassmanian of `planes' through $\mathcal{B}$,
the so-called curvature varifolds.

\item[(\emph{ii})] A generalized notion of curvature can be associated with
curvature varifolds. It enters the constitutive structure of the surface
energy along the crack margins. The resulting energy differs from Griffith's
proposal for the presence of the generalized curvature of the varifold and
the energy along the tip in three dimensions. In this sense the model is an
evolution of Griffith's scheme.
\end{description}

The curvature-dependence of the surface energy has analytical advantages and
permits the control of minimizing sequences. A theorem showing the existence
of pairs of crack patterns and transplacement fields in appropriate measure
and function spaces is then available (see \cite{GMMM}). It has some
implications:

\begin{description}
\item[(\emph{a})] The crack pattern results a rectifiable set. Although it
can be very irregular, it has the features that our intuition assigns to a
fracture.

\item[(\emph{b})] Balance equations can be derived in weak form from the
first variation even for a \emph{generic} rectifiable set. Notice that if a
crack is assumed to be coincident with the discontinuity set of the
transplacement field only, to obtain balance equations, stronger regularity
assumptions on the geometry of the crack pattern are necessary (see detailed
analyses in \cite{BFM}, \cite{BFM2}, \cite{DFT}, \cite{FraMil}, \cite{D}).
\end{description}

These consequences imply that, beyond the analytical advantages, the
physical meaning of items (\emph{i}) and (\emph{ii}) deserves additional
analyses. They can be developed by going along the essential steps of the
theory.

\section{ \ }

Consider the place $\mathcal{B}$ of a body, selected as a reference, as an
open, connected set in the three-dimensional ambient space, with
surface-like boundary oriented by the normal at each point, to within a
finite number of corners and edges. If a $2D$ crack is formed in the
deformed configuration $u\left( \mathcal{B}\right) $ -- $u$ the
transplacement -- and crosses a generic point $u\left( x\right) $ with $x$
in $\mathcal{B}$, its `direction' is locally described by tangent plane to
the crack at $u\left( x\right) $, when the crack is smooth. When the crack
margins have a corner at $u\left( x\right) $, a cone of planes has to be
considered. Crack patterns can be however very irregular. One can accept a
set as a representative of a crack pattern when it is just rectifiable as
mentioned hitherto. So, an approximate notion of tangent plane is available
in geometric measure theory (see \cite{Federer}).

Crack patterns can have a fictitious representation in the reference place $%
\mathcal{B}$ (let say through sets with zero volume measure) -- the
reference place is now mutant because the macroscopic structure of the body
is changing with the nucleation and growth of a crack pattern. Let $\Pi $
indicate a two-dimensional plane or a straight line in three dimensional
ambient space where $\mathcal{B}$ is contained. The pair $\left( x,\Pi
\right) $ gives in $\mathcal{B}$ local information on the geometry of the
crack crossing possibly $u\left( x\right) $. Of course, up to when a real
crack is not realized, any $\Pi $ in the star at $x$ can be a candidate to
describe locally the direction of a possible crack pattern. The pair $\left(
x,\Pi \right) $ can be viewed as a typical point of a fiber bundle ${%
\mathcal{G}}_{k}\left( \mathcal{B}\right) $, $k=1,2$, with natural projector 
$\pi :{\mathcal{G}}_{k}\left( \mathcal{B}\right) \rightarrow \mathcal{B}$
and typical fiber $\pi ^{-1}\left( x\right) ={\mathcal{G}}_{k,3}$, the
Grassmanian of $2D-$planes or straight lines associated with $\mathcal{B}$.
A $k-$\textbf{varifold} over ${\mathcal{B}}$\ is a non-negative Radon
measure $V$ over the bundle ${\mathcal{G}}_{k}({\mathcal{B}})$ (see \cite%
{Almgren}, \cite{Allard1}, \cite{Allard2}, \cite{Hutchinson}, \cite%
{Mantegazza}). The measure $V$ has a projection over ${\mathcal{B}}$
obtained by using the \emph{projector of measures} -- indicated here by $\pi
_{\#}$ -- associated with the natural projection $\pi $ defining the fiber
bundle ${\mathcal{G}}_{k}\left( \mathcal{B}\right) $. Such a projection,
namely $\pi _{\#}V$, is Radon measure over ${\mathcal{B}}$ and is also
indicated for short by $\mu _{V}$. It is called the weighed measure of the
varifold and defines the so-called \emph{mass} $\mathbf{M}\left( V\right) $
of the varifold itself through the relation $\mathbf{M}\left( V\right)
:=V\left( {\mathcal{G}}_{k}({\mathcal{B}})\right) =\mu _{V}\left( {\mathcal{B%
}}\right) $. For the purpose of describing crack patterns through measures,
the essential point is the possibility of constructing varifolds over a
subset $\mathfrak{b}$ of ${\mathcal{B}}$. The subset $\mathfrak{b}$ can have
variegate nature. Here, for the physical purpose at hand, it is assumed that
it is an admissible crack in the sense sketched above. Let $\mathcal{H}^{k}$%
\ be the $k-$dimensional Hausdorff measure in $\mathbb{R}^{3}$, $k=1,2$. It
is assumed that $\mathfrak{b}$ is a $\mathcal{H}^{k}-$measurable, $k-$%
rectifiable subset of ${\mathcal{B}}$. For $\theta $ a function in $%
L^{1}\left( \mathfrak{b},\mathcal{H}^{k}\right) $, the approximate tangent $%
k-$space\footnote{%
The choice $k=1,2$ allows one to treat in a unitary way surface and linear
cracks.} (here $1D$ or $2D$) $T_{x}\mathfrak{b}$ to $\mathfrak{b}$ at almost
every $x$ in $\mathfrak{b}$ is defined for $\theta \mathcal{H}^{k}\llcorner 
\mathfrak{b}$ a.e. $x\in {\mathcal{B}}$. The symbol $\Pi \left( x\right) $
indicates the orthogonal projection onto $T_{x}\mathfrak{b}$. A varifold $V_{%
\mathfrak{b},\theta }$, restricted to $\mathfrak{b}$, can be then defined.
It is called the \textbf{rectifiable varifold} associated with $\mathfrak{b}$%
, with density $\theta $, and is such that%
\begin{equation*}
\int_{{\mathcal{G}}_{k}({\mathcal{B}})}\varphi \left( x,\Pi \right) \text{ }%
dV_{\mathfrak{b},\theta }\left( x,\Pi \right) =\int_{\mathfrak{b}}\theta
\left( x\right) \varphi \left( x,\Pi \right) \text{ }d\mathcal{H}^{k},
\end{equation*}%
for any $\varphi \in C^{0}\left( {\mathcal{G}}_{k}({\mathcal{B}})\right) $.
Rectifiable sets can be considered a sort of generalized surfaces (see \cite%
{Almgren}). A subclass of them admits a notion of generalized mean curvature
vector (see \cite{Allard1}, \cite{Allard2}). For members of such subclass
(not all) a notion of second fundamental form can be defined (see \cite%
{Hutchinson}). Here the attention is on varifolds admitting density $\theta $
with integer values, the so-called \textbf{integer rectifiable varifolds}.
They allow the definition of a special class of varifolds (see \cite%
{Mantegazza}) which is essential for the ensuing developments\footnote{%
See \cite{GiaMart}, last chapter, for a nimble presentation of varifolds.}.

A new ingredient has to be inserted. It is a third-rank tensor field $\left(
x,\Pi \right) \longmapsto \mathbf{A}\left( x,\Pi \right) \in \mathbb{R}%
^{3\ast }\otimes \mathbb{R}^{3}\otimes \mathbb{R}^{3\ast }$ over ${\mathcal{G%
}}_{k}({\mathcal{B}})$, with components $A_{j}^{\ell i}$. It plays the role
of a \textbf{generalized curvature}. A varifold $V$ is called a \textbf{%
curvature }$k-$\textbf{varifold with boundary} if (\emph{i}) $V$ is an
integer, rectifiable $k-$varifold $V_{\mathfrak{b},\theta }$ associated with
the triple $\left( \mathfrak{b},\theta ,\mathcal{H}^{k}\right) $, (\emph{ii}%
) there exists a function $\mathbf{A}\in L^{1}\left( {\mathcal{G}}_{k}({%
\mathcal{B}}),\mathbb{R}^{3\ast }\otimes \mathbb{R}^{3}\otimes \mathbb{R}%
^{3\ast }\right) $,\emph{\ }and a vector Radon measure $\partial V$\emph{\ }%
such that, for every\emph{\ }$\varphi \in C_{c}^{\infty }\left( {\mathcal{G}}%
_{k}({\mathcal{B}})\right) $, one gets%
\begin{equation*}
\int_{{\mathcal{G}}_{k}({\mathcal{B}})}\left( \Pi _{j}^{i}D_{x_{j}}\varphi
+A_{j}^{il}D_{\Pi _{j}^{l}}\varphi +A_{j}^{ij}\varphi \right) \text{ }%
dV\left( x,\Pi \right) =-\int_{{\mathcal{G}}_{k}({\mathcal{B}})}\varphi 
\text{ }d\partial V^{i}\left( x,\Pi \right) .
\end{equation*}%
The vector measure $\partial V$\emph{\ }is called the \textbf{varifold
boundary measure} \cite{Mantegazza}. The subclass of varifolds with
generalized curvature $\mathbf{A}\ $in $L^{p}\left( {\mathcal{G}}_{k}({%
\mathcal{B}})\right) $ is indicated here by $CV_{k}^{p}\left( {\mathcal{B}}%
\right) $. It is possible to show (see \cite{Hutchinson}) that if $V=V_{%
\mathfrak{b},\theta }\in CV_{k}^{p}\left( \mathcal{B}\right) $, with $p>k$, $%
V$ is locally the graph of a multivalued function of class $C^{1,\alpha }$, $%
\alpha =1-\frac{p}{k}$, far from $\partial V$.

\begin{itemize}
\item Varifolds with boundary can be used as descriptors of crack patterns: (%
\emph{i}) The set $\mathfrak{b}$ has the minimal geometrical properties of
an admissible crack, at least in the sense mentioned above. (\emph{ii}) The
density $\theta $ furnishes information on its possible faceted shape%
\footnote{%
If in a neighborhood of $x$ there is a smooth surface, $\theta =1$, when
there is a net fold, $\theta =2$, and so on.}. (\emph{iii}) The local
orientation of the crack pattern is accounted for through $\Pi $. (\emph{iv}%
) Curvature is considered, although in the generalized (weak) form specified
above. (\emph{v}) The boundary of the crack -- it includes the tip -- is
described by the boundary of the varifold.

\item Consider a smooth crack crossing a body in $\mathcal{B}$\ and
intersecting somewhere its boundary but maintaining the tip in the interior
of $\mathcal{B}$. A two-dimensional ($k=2$) varifold $V_{2}$ describes the
crack, its boundary measure $\partial V_{2}$\ is supported by the entire
boundary of the crack itself. To represent separately the crack tip, that is
the part of the boundary of the crack in the interior of $\mathcal{B}$, a
specific one-dimensional varifold $V_{1}$ has to be inserted. It is
supported by the tip alone. Its boundary is supported by possible corners
along the tip and the points determining the intersection of the tip with
the external boundary of the body $\partial \mathcal{B}$. The insertion of $%
V_{1}$ allows one to assign later energy to the tip of the crack. Different
properties can be also assigned to the corners of the tip by using $\partial
V_{1}$. Of course, to capture the intuitive structure of the geometry under
scrutiny, the varifolds $V_{2}$ and $V_{1}$ have to satisfy a certain link.
A definition presented in \cite{GMMM} specifies the link: a family $\left\{
V_{k}\right\} _{k=1}^{d-1}$ of $k-$varifolds with boundary in $d-$%
dimensional ambient space is said to be \textbf{stratified} when $\pi
_{\#}\left\vert \partial V_{k}\right\vert \leq \mu _{V_{k-1}}$ for all $k$%
's. In the special case treated here the condition of stratification reduces
to $\pi _{\#}\left\vert \partial V_{2}\right\vert \leq \mu _{V_{1}}$.

\item The choice $k=1,2$ for constructing ${\mathcal{G}}_{k}({\mathcal{B}})$
and the associated varifolds allows one to consider not only two-dimensional
cracks with the relative tips but also additional linear defects ($k=1$)
which can be cracks included in very thin tubes -- material bonds are broken
along a line for some reason -- or even dislocations. In the latter case,
through a one-dimensional varifold one can describe dislocations emanating
from a crack tip in a three-dimensional body. In the former case one manages
crack patterns stratified over various dimensions. Of course, the analogous
description can be adopted in space dimension $d\geq 2$ and stratification
of defects of various nature can be accounted for.
\end{itemize}

A new form of the energy for a body undergoing fractures, based on the
description of cracks in terms of varifolds, has been proposed in \cite{GMMM}
(see also comments in \cite{2}, \cite{3}). Such an energy is indicated below
by $\mathcal{E}\left( u,\left\{ V_{k}\right\} ,\mathcal{B}\right) $. It
differs from the expression $\mathfrak{E}\left( \mathcal{C},u\right) $\
coming from the traditional Griffith's proposal and is an extension of it.
For a three-dimensional body it reads:%
\begin{eqnarray*}
\mathcal{E}\left( u,\left\{ V_{k}\right\} ,\mathcal{B}\right) &:&=\int_{%
\mathcal{B}}e\left( x,u\left( x\right) ,Du\left( x\right) \right) \text{ }%
dx+\dsum\limits_{k=1}^{2}\alpha _{k}\int_{\mathcal{G}_{k}\left( \mathcal{B}%
\right) }\left\vert \mathbf{A}_{\left( k\right) }\right\vert ^{p_{k}}\text{ }%
dV_{k}+ \\
&&+\dsum\limits_{k=1}^{2}\beta _{k}\mathbf{M}\left( V_{k}\right) +\gamma 
\mathbf{M}\left( \partial V_{1}\right) ,
\end{eqnarray*}%
where $\alpha _{k}$, $\beta _{k}$, $\gamma $ and $p_{k}$\ are constitutive
coefficients. In particular, $\alpha _{k}$, $\beta _{k}$, $\gamma $ are
positive numbers, so the contribution of the generalized curvature of the
varifolds is always present, even if it can be extremely small. The density $%
e\left( x,u,Du\right) $ is defined as the difference $e\left( x,u,Du\right)
:=\tilde{e}\left( x,Du\right) -w\left( u\right) $ between the \textbf{bulk
elastic energy} $\tilde{e}\left( x,Du\right) $ and the \textbf{potential} $%
w\left( u\right) $ \textbf{of external body forces}. A number of comments on
its physical meaning are necessary.

\begin{itemize}
\item The addendum $\beta _{2}\mathbf{M}\left( V_{2}\right) $ has the role
of the last integral in $\mathfrak{E}\left( \mathcal{C},u\right) $, that is
the role of Griffith's surface energy: $\beta _{2}$ has the same meaning of $%
\phi $ in $\mathfrak{E}\left( \mathcal{C},u\right) $. Of course, the first
addendum is the bulk elastic energy with density $e\left( x,Du\right) $ --
the body is then simple but anisotropic -- and coincides with the first
integral in $\mathfrak{E}\left( \mathcal{C},u\right) $. The other terms are
not standard.

\item The addendum $\beta _{1}\mathbf{M}\left( V_{1}\right) $ counts energy
along the tip. Such an energy is proportional to the length of the tip
itself, namely to the mass $\mathbf{M}\left( V_{1}\right) $\ of the
one-dimensional varifold supported by the tip itself.

\item The term $\gamma \mathbf{M}\left( \partial V_{1}\right) $ adds
possible energy concentrated at the tip corners where material bonds can be
entangled in principle in a way different from the other parts of the tip.

\item The two addenda%
\begin{equation*}
\alpha _{2}\int_{\mathcal{G}_{2}\left( \mathcal{B}\right) }\left\vert 
\mathbf{A}_{\left( 2\right) }\right\vert ^{p_{2}}\text{ }dV_{2}+\alpha
_{1}\int_{\mathcal{G}_{1}\left( \mathcal{B}\right) }\left\vert \mathbf{A}%
_{\left( 1\right) }\right\vert ^{p_{1}}\text{ }dV_{1}
\end{equation*}%
mark in a more pronounced manner the difference with respect to $\mathfrak{E}%
\left( \mathcal{C},u\right) $. They have pure configurational nature: they
do not involve directly the gradient of deformation $Du\left( x\right) $.
The first one accounts for the (generalized) curvature of the varifold
describing the surface of the crack, the second one includes the curvature
of the varifold describing the tip. Influence of the curvature of the crack,
above all in the proximity of the tip, has been recognized in \cite{SB}, a
work devoted however to other aspects of fracture processes, namely
Grinfeld's instability. The curvature here aims to account at macroscopic
level of local microstructural effects at low scale, occurring in the
cracking process. Curvature energy can be associated with bending effects in
breaking material bonds when a crack is determined -- consider for example a
material in which the inner bonds are modeled through beam-like
interactions. A point has to be stressed: Bending occurs in the current
configuration $u\left( \mathcal{B}\right) $ while, as a function of $x$, $%
\mathbf{A}$ is defined over the reference configuration $\mathcal{B}$.
However, when bending of material bonds in the actual configuration breaks
the bonds themselves, such a bending has a configurational effect because it
contributes to the mutation of $\mathcal{B}$ due to the nucleation and
possible propagation of a crack. All configurational effects are measured in 
$\mathcal{B}$, as it is commonly accepted. Additionally, analogies with
different approaches to crack analyses in complex bodies (see \cite{JNLS})
and bodies with strain-gradient effects (see \cite{WH}, \cite{RL07}) can be
called upon to enforce the interpretation of the presence of $\mathbf{A}$ as
a configurational indicator of the effects due to latent microstructures.
The terms including $\mathbf{A}$ in the energy tell us essentially that,
once boundary conditions are prescribed, one has to pay in energy for
curving the crack. Consider a rectangular planar sheet of a homogeneous
material half of it including a straight crack in the middle as in Figure 1.
Apply boundary conditions in terms of transplacement (Dirichlet boundary
conditions) in mode 1. If the boundary conditions are such that the crack
can growth, it remains straight, unless some additional agency occurs paying
energy in curving it. \FRAME{ftbpFU}{2.6567in}{1.6855in}{0pt}{\Qcb{Clamped
two-dimensional sheet with a horizontal crack in the middle, subjected to
Dirichlet boundary conditions -- the arrows represent applied
transplacements, the black zone is the clamping device.}}{}{Figure}{\special%
{language "Scientific Word";type "GRAPHIC";maintain-aspect-ratio
TRUE;display "USEDEF";valid_file "T";width 2.6567in;height 1.6855in;depth
0pt;original-width 3.3771in;original-height 2.1309in;cropleft "0";croptop
"1";cropright "1";cropbottom "0";tempfilename
'KMX20T00.wmf';tempfile-properties "XPR";}}

\item Consider a body in an initial configuration in which no crack is
present. For example, $\mathcal{B}$ is a three-dimensional ball. After some
loading program, a certain configuration in which a crack occurs is reached.
Clearly one could consider such a configuration as the initial one. The
possibility stresses the point that `being cracked' is only a relative
concept. Configurations -- or better states -- have to be compared to affirm
that a body is cracked. Moreover, in sequences of configurations (or states)
in which one excludes the possibility of restoring cracks by gluing the
matter across crack facies, an order relation has to be considered. Such an
order is given by monotonicity in crack patterns: if in a certain
configuration there is a crack with respect to an uncracked configuration in
the sequence, a subsequent configuration can have a crack pattern that
coincides with or includes the previous crack. In terms of varifolds such a
point of view is expressed by affirming the existence a family of \emph{%
comparison varifolds }$\left\{ \tilde{V}_{k}\right\} $ such that the family
of varifolds $\left\{ V_{k}\right\} $ describing the actual crack pattern is
constrained by $\mu _{\tilde{V}_{k}}\leq \mu _{V_{k}}$ for any $k$ and $%
\tilde{V}_{k}\in CV_{k}^{p_{k}}\left( \mathcal{B}\right) $. The assignment
of $\left\{ \tilde{V}_{k}\right\} $ does not mean that one is considering in
a given configuration a pre-existing crack pattern always, because the
comparison varifold family can be also empty.

\item The assumption used so far that the ambient space is three-dimensional
has been accepted only to be close to the standard physical intuition. There
is no obstruction to consider higher dimensional spaces. Consider the
dimension of the ambient space to be $d\geq 2$. The previous treatment can
be extended to this case straight away. The only modification, including the
expression of the energy is that the range of $k$ must be considered to be $%
1\leq k\leq d-1$, at least in principle. The rest remains unchanged. This
choice allows one to consider stratified families of defects at various
scales with various physical meanings, depending on the circumstance. Of
course, there could be cases in which $k$ could not take all the natural
values from $1$ to $d-1$. Even in three-dimensional ambient space one could
have only $k=1$, for example, being in the condition to consider only linear
defects like nets of dislocations. Such a case would deserve perhaps a
treatment a part. The case of four manifolds, $d=4$, could be called upon
when the description of crack in relativistic elastic bodies is considered,
with all the necessary changes in the representation of the energy, adapted
to the relativistic setting. In this case, however, care must be taken
because in the relativistic setting a representation of continuous bodies
based on the back-to-label representation seems to be preferable. The
circumstance then would change the stage and would include a sort of `mixed'
representation of the energy.
\end{itemize}

In the setting described so far, a minimality requirement is prescribed for
the energy:

\begin{center}
\emph{Minimize }$E\left( u,\left\{ V_{k}\right\} ,\mathcal{B}\right) $\emph{%
\ with }$V_{k}$\emph{\ in }$CV_{k}^{p_{k}}\left( \mathcal{B}\right) $\emph{,
comparison varifolds }$\left\{ \tilde{V}_{k}\right\} $\emph{, }$u$\emph{\ in
an appropriate function space, with assigned boundary conditions. }
\end{center}

Solution to this problem, if any, is a pair $\left( u,\left\{ V_{k}\right\}
\right) $. Minimization over a class of varifolds has the meaning of
minimization over a class of bodies: every possible crack pattern
represented over $\mathcal{B}$ (remind that nucleation and\TEXTsymbol{%
\backslash}or growth of cracks occur in the current place) defines a body.
Different crack patterns indicate different bodies: $\mathcal{B}$, in fact,
changes. The family of varifolds $\left\{ V_{k}\right\} $ represents the
crack pattern -- so it selects a body -- and the field $x\longmapsto u\left(
x\right) $ describes the deformation of such a body. The two `objects' are
correlated. In fact, in the actual configuration possible nucleation, growth
and/or opening of a crack are consequences of the deformation, the varifolds
over $\mathcal{B}$\ are\ representatives of what happens in the actual
configuration. The choice of the function space hosting the generic $u$ is
then another key point of the treatment. It has to be linked to the
varifolds supported on $\mathcal{B}$.

\section{ \ \ }

In continuum mechanics the choice of function spaces as ambient for
solutions to equilibrium or evolution problems has constitutive nature. The
properties of the members of a given space carry a physical meaning about.
The characteristic physical features of the problem under analysis have then
to address the functional choice. In the case under scrutiny, the idea is
that the material is elastic-brittle. It means that the material is elastic
up to a certain threshold after which a crack is created while outside the
crack the material is still in elastic phase. The threshold can be expressed
in terms of deformation, stress, energy, depending on circumstances.
Moreover, as it will be clear later, the threshold can be also not expressed
directly. It is, in a sense, included in the choice of the function space
that one selects.

Let us focus the attention on pure elasticity first. If the material is
purely elastic, it can deform at will, without any threshold and, in
principle, without end. The deformation can be also perfectly recovered,
after unloading. In this sense, phenomena like cavitation in solids are
ascribed to elastic-brittle behavior rather than perfect elastic setting. If
this view is accepted, the consequence would be that a perfectly
(hyper)elastic body should be such that any compatible transplacement (or
displacement, depending on the choice) field does not describe the
nucleation of fractures or holes.

It is well known that under conditions of polyconvexity of the elastic
energy density, minimizers of the elastic energy of a simple body can be
found in the Sobolev space $W^{1,p}\left( \mathcal{B},\mathbb{R}^{3}\right) $
-- a space hosting maps with first distributional derivative having
integrable $p-$power, i.e. the first derivative is in $L^{p}$. However, when 
$p<3$ non negligible is the presence of transplacement maps with graphs
admitting boundaries with projections into the interior of $\mathcal{B}$.
Such boundaries describe the formation of `holes' and/or open `fractures' of
various nature, so they are undesirable when a purely elastic material is
under analysis, at least if the view on elasticity sketched above is
accepted.

The difficulty can be overcame. In fact, there is a global way to check --
eventually to control -- the presence of boundaries in the graph of a map
with projection into the domain of the map itself through linear
functionals. Indicate them by $G_{u}$, with the indices $u$ suggesting that
the functional $G$ is associated with the transplacement $u$. In the case
treated here, such functionals are linear over smooth $3-$forms with compact
support in $\mathcal{B}\times \mathbb{R}^{3}$. The general technique, which
is valid in any finite dimension $d$ of the ambient space, is described in
the monograph \cite{GMS}. Here, I sketch only minimal ideas, furnishing the
essential picture and physical interpretations in the case under scrutiny.

Preliminarily, remind that a $r-$vector over a linear space $E$ is a rank$-r$
skew-symmetric tensor, that is an element of the skew-symmetrization of $E%
\overset{r}{\overbrace{\otimes ...\otimes }}E$. The space of $r-$vectors is
indicated here by $\Lambda _{r}\left( E\right) $. It has a natural dual $%
\Lambda ^{r}\left( E\right) $. Any map of the type $\omega :\mathcal{B}%
\longrightarrow \Lambda ^{r}\left( E\right) $ is called a $r-$form\footnote{%
More specifically, one should consider $E$ as a real vector bundle over $%
\mathcal{B}$ of fiber dimension $d$, that is a family of $d-$dimensional
vector spaces parametrized by points of $\mathcal{B}$.}. The space of all $%
r- $forms of the type just defined is indicated by $\mathcal{D}^{r}\left(
E\right) $

For the mechanics treated here interesting is the case of $3-$vectors over $%
\mathbb{R}^{3}\times \mathbb{R}^{3}$, the space hosting the graph of the
deformation. Consider a deformation (transplacement) $x\longmapsto u\left(
x\right) \in \mathbb{R}^{3}$, $x\in \mathcal{B}$. Amid all possible $3-$%
vectors over $\mathbb{R}^{3}\times \mathbb{R}^{3}$, it is possible to define
at every $x\in \mathcal{B}$ the $3-$vector $M\left( Du\right) $ associated
with the gradient of deformation $Du$. Detailed definition and various
properties can be found in \cite{GMS}, here, for the expository purposes
declared at the beginning, it is only necessary to know that at each $x$ its
components are the entries of $Du\left( x\right) $, $adjDu\left( x\right) $, 
$detDu\left( x\right) $. In $M\left( Du\right) $, then, all elements
characterizing the deformation of lines, areas, and volume of the body in $%
\mathcal{B}$ are included. In this sense, $M\left( Du\right) $ characterizes
completely the deformation. Its dual counterpart -- the value at the same $x$
of some form in $\mathcal{D}^{3}\left( \mathbb{R}^{3}\times \mathbb{R}%
^{3}\right) $ -- is then a sort of generalized stress.

Given a transplacement $x\longmapsto u\left( x\right) \in \mathbb{R}^{3}$, $%
x\in \mathcal{B}$, the $3-$\textbf{current integration} $G_{u}$ (\textbf{%
current} for short) over the graph of $u$ is defined to be a linear
functional over smooth $3-$forms $\omega \in \mathcal{D}^{3}\left( \mathbb{R}%
^{3}\times \mathbb{R}^{3}\right) $ with compact support in $\mathcal{B}%
\times \mathbb{R}^{3}$, namely%
\begin{equation*}
G_{u}\left( \omega \right) :=\int_{\mathcal{B}}\left\langle \omega \left(
x,u\left( x\right) \right) ,M\left( Du\left( x\right) \right) \right\rangle 
\text{ }dx\mathbf{,}
\end{equation*}%
where the angle brackets indicate the natural action over $M\left( Du\right) 
$ of its dual counterpart\footnote{%
A bit more precisely, the current is defined by taking the rectifiable part
of the graph of $u$, that is the part of the graph that can be seen as the
graph of Lipshitz maps.}. Essentially, $G_{u}\left( \omega \right) $\ plays
the role of generalized internal power. The number $\mathbf{M}\left(
G_{u}\right) $ indicates here the so-called \textbf{mass of the current} and
is defined by%
\begin{equation*}
\mathbf{M}\left( G_{u}\right) :=\int_{\mathcal{B}}\left\vert M\left(
Du\left( x\right) \right) \right\vert \text{ }dx,
\end{equation*}%
where $\left\vert M\left( Du\left( x\right) \right) \right\vert $ is the
modulus of $M\left( Du\left( x\right) \right) $, evaluated in the standard
way for tensors. The symbol $\left\vert G_{u}\right\vert $ indicates the 
\textbf{total variation }of the current and is defined as usual for
functionals.\ A \textbf{boundary current} can be associated with $G_{u}$: it
is indicated by $\partial G_{u}$ and defined by duality, that is\footnote{%
The so-called external differentiation over forms is indicated by $d$ and
acts as $d:\mathcal{D}^{n}\left( E\right) \rightarrow \mathcal{D}%
^{n+1}\left( E\right) $.} 
\begin{equation*}
\partial G_{u}\left( \omega \right) :=G_{u}\left( d\omega \right) ,\ \forall
\omega \in \mathcal{D}^{2}(\mathcal{B}\times \mathbb{R}^{3}),
\end{equation*}%
with $\mathcal{D}^{2}(\mathcal{B}\times \mathbb{R}^{3})$ the space of $2-$%
forms over $\mathbb{R}^{3}\times \mathbb{R}^{3}$\ with compact support%
\footnote{%
The definitions of currents and related boundaries can be also available in
spaces with higher dimension (see \cite{GMS} for the complete theory).} in $%
\mathcal{B}\times \mathbb{R}^{3}$. The notion of boundary current has not
only formal nature. It has an immediate physical interpretation: when the
graph of $u$ is free of boundaries inside the interior of $\mathcal{B}$, $%
\partial G_{u}\left( \omega \right) =0$ for any $\omega \in \mathcal{D}^{2}(%
\mathcal{B}\times \mathbb{R}^{3})$. Essentially, this zero boundary
condition prevents the formation of cracks or holes inside the actual place $%
u\left( \mathcal{B}\right) $ of the body. Such a condition has been used
(see \cite{GMS} and the other references of its authors mentioned therein)
to define a class of transplacements -- the so-called \textbf{weak
diffeomorphisms}\emph{\ }-- which is `constitutively' an appropriate choice
for describing what one imagines to be a pure elastic deformation, as
sketched above.

Here the situation is a bit more complicated. The deformation has to be a
weak diffeomorphism outside a subset of the support of the varifolds
describing the crack pattern, if the minimizing procedure provides a
non-empty family of minimizing varifolds. Inside that subset, the
transplacement admits jumps, so it is not purely a weak diffeomorphism.

Extended weak diffeomorphisms are then necessary. They must have the
physical properties just indicated. Such properties can be summarized in a
formal definition.

\begin{defn}
Assigned a stratified curvature varifold $V=\left\{ V_{k}\right\}
_{k=1}^{n-1}$ with boundary, i.e., $V_{k}\in CV^{p_{k}}$, a map $%
x\longmapsto u$ is said to be an \textbf{extended} \textbf{weak
diffeomorphism} (in short $u\in dif^{1,1}(\mathcal{B},V,\mathbb{R}^{3})$),
when
\end{defn}

\begin{description}
\item[(\emph{i})] $u\in L^{1}\left( \mathcal{B}\right) $ \emph{and is a.e.
approximately differentiable,}

\item[(\emph{ii})] $\left\vert M\left( Du\right) \right\vert \in L^{1}\left( 
\mathcal{B}\right) $\emph{,}

\item[(\emph{iii})] $\det Du\left( x\right) >0$ \ \ \emph{for almost every} $%
x\in \mathcal{B}$\emph{,}

\item[(\emph{iv})] \emph{for any} $f\in C_{c}^{\infty }(\mathcal{B}\times 
\mathbb{R}^{3})$
\end{description}

\begin{equation*}
\int_{\mathcal{B}}f\left( x\mathbf{,}u\left( x\right) \right) \det Du\left(
x\right) \text{ }dx\leq \int_{\mathbb{R}^{3}}\sup_{x\mathbf{\in }\mathcal{B}%
}f\left( x,w\right) \text{ }dw\mathbf{,}
\end{equation*}

\begin{description}
\item[(\emph{v})] $\pi _{\#}\left\vert \partial G_{u}\right\vert \leq
\dsum\limits_{j=1}^{2}\mu _{V_{k}}+$ $\pi _{\#}\left\vert \partial
V_{1}\right\vert $ \ \emph{as measures on }$\mathcal{B}$\emph{.}
\end{description}

The definition has natural extension in $\mathbb{R}^{d}$: the summation in
the item (\emph{v}) should be extended up to $d-1$. Of course, the
definition of currents and related boundaries holds in dimension $d$: in
that case $M\left( Du\right) $ is tested over $d-$forms.\ Here and in the
whole paper, the restriction $d=3$ is essentially motivated by the physics
under scrutiny (see \cite{GMMM} for the abstract theory). The first item
indicates the possibility of evaluating the gradient of deformation. The
second item is another regularity condition. It implies that one can in
principle measure the average of the gradient of deformation, the volume
change, the overall deformation of surfaces. The third item is the standard
condition that a transplacement be an orientation preserving map. Item (%
\emph{iv}) is the condition mentioned at the beginning of Section 1. It
permits to move $\mathcal{B}$ along $u$\ into a region $u\left( \mathcal{B}%
\right) $ in such a way that self-contact between parts of the boundary $%
\partial \mathcal{B}$ be allowed while self-penetration excluded. Notice
that in item (\emph{v}) the action of the projector $\pi _{\#}$ on the total
variation of the boundary current is motivated by the fact that the latter
behaves substantially as a measure. Essential properties for the space of
extended weak diffeomorphisms are shown in \cite{GMMM} (see there the
relevant theorems and proofs).

Standard weak diffeomorphisms are $W^{1,1}\left( \mathcal{B},\mathbb{R}%
^{3}\right) $ maps that satisfy the items (\emph{ii}), (\emph{iii}), (\emph{%
iv}) in previous definition while item (\emph{v}) which is substituted by
the zero boundary condition $\partial G_{u}\left( \omega \right) =0$ for any 
$\omega \in \mathcal{D}^{2}(\mathcal{B}\times \mathbb{R}^{3})$. From a
kinematic point of view, in going from $dif^{1,1}(\mathcal{B},\mathbb{R}%
^{3}) $ to $dif^{1,1}(\mathcal{B},V,\mathbb{R}^{3})$, one transits from pure
elastic setting to elastic-brittle behavior.

In this sense, the requirement of minimality of the energy includes the
possibility of finding minimizers -- if any -- in terms of weak
diffeomorphisms and null varifolds, and in terms of extended weak
diffeomorphisms and non-null varifolds. The transition from a situation to
another is morally the threshold from the elastic to the elastic-brittle
behavior. In this sense, also, there is no need in principle of adding
another condition defining the threshold itself. There could be also
minimizers for which the transplacement field is simply a weak
diffeomorphism but the varifolds are not null. This situation describes
presence of closed cracks only: in other words, material bonds are broken
but the crack remains closed and the transplacement field does not jump
across the crack facies. Such a situation can occur in a step-by-step
minimization program obtained by updating in time steps the boundary
conditions and requiring minimality of the energy at each step. At the step $%
n-1$ a crack pattern can occur, at the step $n$ the deformation closes the
cracks, at the step $n+1$ there is a purely elastic continuation, then, at
further steps, new crack patterns accrue.

Of course, proving existence of minimizers is a crucial step for attributing
sense to the previous reasonings. Existence depends on the characteristic
properties of the energy and the boundary conditions. Once the existence of
minimizers is established, the characterization of them along the physical
suggestions collected above is matter of regularity theorems. The question
is open. Actually, any regularity theorem is available. Different is the
case of the existence problem.

\section{ \ }

The existence result for the minimum problem stated above for the energy $%
\mathcal{E}\left( u,\left\{ V_{k}\right\} ,\mathcal{B}\right) $ of an
elastic brittle solids has been proven in \cite{GMMM}. Boundary conditions
of Dirichlet type can be presumed. They are given by prescribing the
transplacement field along the boundary $\partial \mathcal{B}$ of $\mathcal{B%
}$.

The discussion of the existence is developed by taking first a subspace of $%
dif^{1,1}(\mathcal{B},V,\mathbb{R}^{3})$, precisely the space $dif^{p,1}(%
\mathcal{B},V,\mathbb{R}^{3})$ defined by%
\begin{equation*}
dif^{p,1}(\mathcal{B},V,\mathbb{R}^{3}):=\left\{ u\in dif^{1,1}(\mathcal{B}%
,V,\mathbb{\hat{R}}^{3})\text{ }|\text{ }\left\vert M\left( Du\right)
\right\vert \in L^{p}\left( \mathcal{B}\right) \right\} ,
\end{equation*}%
for some $p>1$. Essentially, the choice of $dif^{p,1}(\mathcal{B},V,\mathbb{R%
}^{3})$ with $p>1$ is a request of additional regularity which is sometimes
necessary for physical needs. Combination with the space of varifolds allows
one to recognize a natural ambient in which the existence of minimizers of
the energy $\mathcal{E}\left( u,\left\{ V_{k}\right\} ,\mathcal{B}\right) $
can be investigated. Such a space is indicated by $\mathcal{A}%
_{q,p,K,\left\{ \tilde{V}_{k}\right\} }\left( \mathcal{B}\right) $ and
defined by%
\begin{eqnarray*}
\mathcal{A}_{q,p,K,\left\{ \tilde{V}_{k}\right\} }\left( \mathcal{B}\right)
&:&=\left\{ \left( u,\left\{ V_{k}\right\} \right) \text{ }|\text{ }V_{k}\in
CV_{k}^{p_{k}}\left( \mathcal{B}\right) ,u\in dif^{q,1}(\mathcal{B},V_{k},%
\mathbb{R}^{3}),\right. \\
&&\left. \left\{ V_{k}\right\} \text{ is stratified},\left\Vert u\right\Vert
_{L^{\infty }\left( \mathcal{B}\right) }\leq K,\mu _{\tilde{V}_{k}}\leq \mu
_{V_{k}},\forall k=1,2\right\} ,
\end{eqnarray*}%
where $\tilde{V}_{1}$ and $\tilde{V}_{2}$\ are comparison varifolds
describing possible initial cracks. In particular, the subspace%
\begin{equation*}
\mathcal{A}_{q,p,K,\left\{ \tilde{V}_{k}\right\} }^{u_{0}}\left( \mathcal{B}%
\right) :=\left\{ \left( u,\left\{ V_{k}\right\} \right) \in \mathcal{A}%
_{q,p,K,\left\{ \tilde{V}_{k}\right\} }\left( \mathcal{B}\right) \text{ }|%
\text{ }u\left( x\right) =u_{0}\left( x\right) ,x\in \partial \mathcal{B}%
_{u},\right\}
\end{equation*}%
with $\partial \mathcal{B}_{u}$ the part of the boundary of the body where
the transplacement field is prescribed, takes into account the boundary
conditions of Dirichlet type mentioned above.

In all these definitions, another regularity requirement is prescribed. In
fact, the condition $\left\Vert u\right\Vert _{L^{\infty }\left( \mathcal{B}%
\right) }\leq K$ imposes that the essential supremum of $u$ is almost
everywhere -- with respect to the Lebesgue measure -- bounded. In fact, a
priori it is not possible to exclude that, if one is able to prove under
some conditions the minimality of the energy $\mathcal{E}\left( u,\left\{
V_{k}\right\} ,\mathcal{B}\right) $ over some space of extended weak
diffeomorphisms and varifolds, the minimizing varifold does not describe a
fragmentation of the body -- let say a crack cutting a piece of matter from
the rest. In this case, a transplacement field could be such that the cut
piece, now free from boundary conditions, can be translated rigidly to
infinity. By imposing that $\left\Vert u\right\Vert _{L^{\infty }\left( 
\mathcal{B}\right) }\leq K$, with $K$ a real number, then, one wants to
avoid the situation just sketched. So, in this sense the assignment of $K$
has not properly constitutive nature. It is not related to some property of
the material, rather it is a parameter selecting admissible deformation
processes, admissibility considered with reference to the possible
unconstrained extraction of pieces of matter from the body.

For $d$ the dimension of the ambient space and $k$ ranging from $1$ to $d-1$%
, analogous definitions of $\mathcal{A}_{q,p,K,\left\{ \tilde{V}_{k}\right\}
}^{u_{0}}\left( \mathcal{B}\right) $ hold and the theory can be generalized
(see relevant results in \cite{GMMM}).

Another crucial point in the path leading to the proof of existence theorem
of minimizers of the energy is the discussion of the structural properties
of the energy. They have constitutive nature, of course.

In non-linear elasticity of simple Cauchy's bodies, common assumptions about
the structure of the energy $e\left( x,u\left( x\right) ,Du\left( x\right)
\right) $ are well known. By indicating by $M_{3\times 3}^{+}$\ the space of 
$3\times 3$ matrices with positive determinant, the energy density $e$ is
considered as a map%
\begin{equation*}
e:\mathcal{B}\times \mathbb{R}^{3}\times M_{3\times 3}^{+}\rightarrow \left[
0,+\infty \right]
\end{equation*}%
with values $e\left( x,u\left( x\right) ,Du\left( x\right) \right) $. Remind
that positiveness of the determinant of $Du\left( x\right) $ is the
condition assuring that the transplacement be orientation preserving. The
properties H1-H4 below are then assumed to hold.

\begin{description}
\item[H1] $e:\mathcal{B}\times \mathbb{R}^{3}\times M_{3\times
3}^{+}\rightarrow \left[ 0,+\infty \right] $ is \textbf{continuous} in $%
\left( x,u\right) $.

\item[H2] The map $Du\left( x\right) \longmapsto e\left( x,u\left( x\right)
,Du\left( x\right) \right) $ is \textbf{polyconvex}: that is there exists a
Borel function $Pe$ acting as%
\begin{equation*}
Pe:\mathcal{B}\times \mathbb{R}^{3}\times \Lambda _{3}(\mathbb{R}^{3}\times 
\mathbb{R}^{3})\rightarrow \mathbb{\bar{R}}^{+},
\end{equation*}%
with values $Pe\left( x,u\left( x\right) ,\xi \left( x\right) \right) $,
which is continuous in $\left( x,u\right) $ for every $\xi \in \Lambda _{3}(%
\mathbb{R}^{3}\times \mathbb{R}^{3})$, convex and lower semicontinuous in $%
\xi $ for every $\left( x,u\right) $, and such that\footnote{%
The dependence of $u$ and $Du$ on $x$ is now suppressed for the sake of
brevity.} $Pe\left( x,u,M\left( Du\right) \right) =e\left( x,u,Du\right) $
for any list of entries $\left( x,u,Du\right) \in \mathcal{B}\times \mathbb{R%
}^{3}\times M_{3\times 3}^{+}$ with $\det Du>0$.

\item[H3] The energy density $e$ satisfies the growth condition%
\begin{equation*}
e\left( x,u,Du\right) \geq C_{1}\left\vert M\left( Du\right) \right\vert
^{r}.
\end{equation*}

\item[H4] For every $x\in \mathcal{B}$ and $Du\in M_{3\times 3}^{+}$, if for
some $u\in \mathbb{R}^{3}$ the inequality $e\left( x,u,Du\right) <+\infty $
is satisfied, then $\det Du>0$.
\end{description}

The physical nature of these assumptions is discussed in various treatises
(see \cite{MH}, \cite{Sil}). The standard presence in the polyconvex energy
of the determinant of the gradient of deformation and the relevant adjugate
is summarized here in the functional dependence on $M\left( Du\right) $. The
choice is not only formal. It furnishes a rapid path toward the extension of
the treatment to $d-$dimensional cases (see \cite{GMS}).

However, the essential point is to underline that, in the setting explored
here, H1-H4 do not need to be supplemented by additional structural
assumptions on the energy to assure the existence of minimizers of $\mathcal{%
E}\left( u,\left\{ V_{k}\right\} ,\mathcal{B}\right) $. The relevant theorem
reads as follows:

\begin{thm}[\protect\cite{GMMM}]
Assume $K>0,$ $q,p_{k}>1$, and $\tilde{V}_{k}\in CV_{k}^{p_{k}}\left( 
\mathcal{B}\right) $\ for any $k$. If there exists $\left( u_{0},\left\{
V_{k}^{0}\right\} \right) \in \mathcal{A}_{q,p,K,\left\{ \tilde{V}%
_{k}\right\} }^{u_{0}}\left( \mathcal{B}\right) $ such that $\mathcal{E}%
\left( u_{0},\left\{ V_{k}^{0}\right\} ,\mathcal{B}\right) <+\infty $, then $%
\mathcal{E}\left( u,\left\{ V_{k}\right\} ,\mathcal{B}\right) $ attains in
that space the minimum value.
\end{thm}

Proof is presented in \cite{GMMM}. Here just comments have to be added.

\begin{itemize}
\item Since no additional structural hypotheses besides H1-H4 of standard
non-linear elasticity need to be added, the information about the possible
nucleation of a crack or growth of an existing one is furnished by the
presence of the terms ruled by varifolds in the energy and the \emph{%
constitutive} choice of $\mathcal{A}_{q,p,K,\left\{ \tilde{V}_{k}\right\}
}^{u_{0}}\left( \mathcal{B}\right) $ as functional setting. It is just the
latter choice that avoids the introduction of an external criterion for the
nucleation of a crack or the growth of an existing one. In fact, the energy
is minimized over a class of possible bodies.

\item More in general than other descriptions, it is possible to determine
the weak form of balance equations for crack patterns which are just
rectifiable sets. The result is not discussed here for the sake of
conciseness. It is presented in \cite{GMMM} and opens the way to
computational opportunities not explored yet. The balance equations derived
naturally in \cite{GMMM}\ for very general crack geometries -- as mentioned
above the crack pattern has to be just a rectifiable set -- are the ones
obtained by horizontal variations, that are variations of the reference place%
\footnote{%
See \cite{Eshelby} for clear explanations on the connection between
horizontal variations describing the potential movement of defects and the
balance of configurational forces. Take into account also that there is
basic difference between the balance of configurational actions associated
with macroscopic mutations of the reference place and the balances of
standard actions generated by the deformation. In the conservative case and
with reference to smooth fields, the former balances are essentially the
pull-back in the reference place of the latter balances. In general it is
not so. The difference has been evidenced first in \cite{GH} vol.1, pages
152-153.}. Thus they have configurational nature: they involve in fact the
Hamilton-Eshelby tensor and non-standard terms deriving from the variations
of the terms including the varifolds, the ones directly related with the
geometry of the crack pattern.

\item The existence result holds also at dimensions greater than 3, provided
that the obvious variations in previous definitions (see \cite{GMMM}, \cite%
{2} for the abstract theory).

\item An analogous result holds also for the generalized energy%
\begin{eqnarray*}
\mathcal{E}\left( u,\left\{ V_{k}\right\} ,\mathcal{B}\right) &:&=\int_{%
\mathcal{B}}e\left( x,u\left( x\right) ,Du\left( x\right) \right) \text{ }%
dx+\dsum\limits_{k=1}^{d-1}\alpha _{k}\int_{\mathcal{G}_{k}\left( \mathcal{B}%
\right) }\phi _{k}\left( \left\vert \mathbf{A}_{\left( k\right) }\right\vert
\right) \text{ }dV_{k}+ \\
&&+\dsum\limits_{k=1}^{d-1}\beta _{k}\mathbf{M}\left( V_{k}\right) +\gamma 
\mathbf{M}\left( \partial V_{1}\right) ,
\end{eqnarray*}%
with $\phi _{k}\left( \mathbf{\cdot }\right) $ a convex real-valued function
satisfying the condition $\phi _{k}\left( t\right) \geq ct^{p_{k}}$. The
interest of this remark is not only technical. It gives a grater degree of
freedom in selecting further constitutive structures under the suggestions
of possible experimental evidences and numerical tests.

\item The existence result does not exclude the possibility that the
minimization procedure foresees stratified varifolds supported on the
boundary of $\mathcal{B}$. In this case one can say that a boundary crack
appears. The meaning of such a boundary cracks is not exotic. In fact, on a
part of the boundary where the transplacement is imposed a crack can occur
so that the boundary condition is `broken'. Consider for example a beam
jointed at one of its ends. Boundary conditions and properties of the
material can be such that a crack occurs just at the interface between the
joint and the beam. However, in principle a boundary crack can appear on a
free part of the boundary. Such a situation can describe the fragmentation
of a thin film at the boundary, which is, essentially, the abrasion of the
boundary itself.
\end{itemize}

\section{ \ \ }

The technique discussed previously is a general tool for the description of
phenomena in which energy can in principle be concentrated over submanifolds
of a certain manifold, and this energy depends on the geometry of the
submanifold itself. It can be used to analyze either specific situations of
physical interest or to formulate and analyze abstract mathematical problems.

\begin{itemize}
\item Phenomena of physical interest that can be described by using the
tools mentioned hitherto deal for example with the mechanics of linear
defects like dislocations and discontinuity surfaces. For dislocations the
choice of the varifolds play a crucial role. For interfaces, the functional
setting has to be changed and a new special class of extended weak
diffeomorphisms arises. The new choice requires proof of completeness of the
new functional class and evaluation of the applicability of lower
semicontinuity results.

\item Another point of discussion is also the evaluation of the crack
nucleation and growth in complex bodies. The adjective `complex'
distinguishes bodies characterized by a prominent influence of changes in
material texture (the \textbf{microstructure}) on the macroscopic behavior,
an influence exerted through inner actions requiring a representation going
beyond the common picture in terms of standard stresses. Quasicrystals,
ferroelectrics, magnetoelastic materials, polymeric bodies of various
nature, including elastomers, fullerene-based composites, porous bodies,
bodies with continuous distributions of dislocations, multiphase materials
are paradigmatic examples. Notwithstanding the variety of special models, a
unitary picture of the mechanics of complex bodies exists. Within it, the
representation of bodies goes beyond standard Cauchy's approach in the sense
that every material element is viewed as a system rather than a black box
individuated by a single point in the ambient space, which is Cauchy's view.
Such a description is multifield and intrinsically multiscale. A \textbf{%
morphological descriptor} field $x\longmapsto \nu \left( x\right) $, $x\in 
\mathcal{B}$, of the essential geometrical features of the material
microstructure is then introduced. To construct the essential structures of
the relevant mechanics, it is just necessary to presume that $\nu \left(
x\right) $ is an element of a set $\mathcal{M}$ which has just the structure
of a differentiable manifold. It is assumed to be finite-dimensional for the
sake of simplicity. $\mathcal{M}$ is called the \textbf{manifold of
substructural shapes}. The interest of mentioning here complex bodies is
motivated by data showing that the microstructural changes may influence in
non negligible way the force driving crack tips along evolution processes.
Theoretical analysis of this phenomenon within the setting of the general
model building framework of the mechanics of complex bodies (that is without
specifying the type of microstructure) has been developed in \cite{JNLS}%
\footnote{%
There one can find appropriate references to works presenting and discussing
experimental data.} from a point of view different from the one adopted
here. The simplest extension of the theory discussed here to complex bodies
is given by an energy of the type%
\begin{eqnarray*}
\mathcal{E}\left( u,\left\{ V_{k}\right\} ,\mathcal{B}\right) &:&=\int_{%
\mathcal{B}}e\left( x,u\left( x\right) ,\nu \left( x\right) ,Du\left(
x\right) ,D\nu \left( x\right) \right) \text{ }dx+ \\
&&+\dsum\limits_{k=1}^{2}\alpha _{k}\int_{\mathcal{G}_{k}\left( \mathcal{B}%
\right) }\left\vert \mathbf{A}_{\left( k\right) }\right\vert ^{p_{k}}\text{ }%
dV_{k}+\dsum\limits_{k=1}^{2}\beta _{k}\mathbf{M}\left( V_{k}\right) +\gamma 
\mathbf{M}\left( \partial V_{1}\right) ,
\end{eqnarray*}%
with the natural modifications in generic dimension $d$. In analyzing the
existence of minimizers, an essential point is the choice of the space
hosting the morphological descriptor maps. Such a choice could require the
embedding of $\mathcal{M}$ into a linear space. Such embedding always exists
because $\mathcal{M}$ is finite dimensional, also it can be isometric when $%
\mathcal{M}$ is Riemannian. In all cases, however, it is not unique so that
it becomes a ingredient of the model, a sort of constitutive choice. Another
point is the link of the jump set of the morphological descriptor field with
the varifolds. Here the underlying physics is subtle. In principle one can
accept that $x\longmapsto \nu \left( x\right) $ may have jumps even outside
the support of the varifolds and there it may be even continuous. Jumps
outside the varifolds can be justified by the formation of domains of
microstructures, like polarization or magnetization domains. The meaning of
the possible continuity on the support of the varifolds is associated with
the question whether in cracking a body one alters along the margins of the
crack the microstructure, in a sense determining a new type of
microstructure, or, else, the microstructure remains the same across the
margins of the crack. In the philosophy of continuum mechanics, the remark
above coincides with asking whether a crack just divides neighboring
material elements or breaks the material elements met in front of the tip.
The answer cannot be definitive and is matter of modelling. The situation
becomes also more complicated when the structure of the surface energy
involving the generalized curvature of the varifold is enriched by making
more articulated assumptions. Relevant investigations are actually open.

\item A point which may deserve to be noted is that the scheme discussed in
previous sections has intrinsic similarity with the general framework of the
mechanics of complex bodies which has been sketched rapidly in the last
item. In fact, when curvature $k-$varifolds with boundary are chosen to
represent cracks, in principle the region where they localize is not known
-- in other words one does not know where the support of the varifolds is
placed in $\mathcal{B}$, that is where the crack is. In this setting, every
point can be crossed in principle by a crack. The minimization procedure
tells us that the crack is here or there, before nothing is known about its
position. Instead of assigning to each material element a morphological
descriptor $\nu $ selected in a finite dimensional differentiable manifold $%
\mathcal{M}$, \emph{one is then assigning to each material element a measure}%
. In this sense, the scheme discussed here is driven by the ideas of the
mechanics of complex bodies, and, in some sense, it goes beyond them a bit.
\end{itemize}

\ \ \ \ \ \ \ 

\textbf{Acknowledgements}. Discussions with Mariano Giaquinta, Giuseppe
Modica, Gianfranco Capriz, Gilles Francfort, Luigi Ambrosio have pushed me
in various manners to write this paper. I thank them for the impulse. This
work has been developed within the activities of the research group in
`Theoretical Mechanics' of the `Centro di Ricerca Matematica Ennio De
Giorgi' of the Scuola Normale Superiore at Pisa. The support of the
GNFM-INDAM is acknowledged.

\end{document}